\documentstyle[12pt]{article}

\newcommand{\tr}{\mbox{tr}\:}
\newcommand{\pa}{\partial}
\newcommand{\la}{\lambda}
\newcommand{\bit}{\begin{itemize}}
\newcommand{\eit}{\end{itemize}}
\newcommand{\ben}{\begin{enumerate}}
\newcommand{\een}{\end{enumerate}}
\newtheorem{teo}{Theorem:}
\title{Separation of variables via integral transformations }
\author{Yaroslav V. Lisitsyn${}\sp a$, Alexander V. Shapovalov${}\sp b$}
\date{}
\begin{document}
\maketitle

{\it
${}\sp a$ Tomsk State University, Tomsk, Russia\\
e-mail: lisitsyn@tspi.tomsk.su

${}\sp b$ Tomsk State University, Tomsk, Russia\\
e-mail: shpv@phys.tsu.tomsk.su}

\begin{abstract}
For  a
system of linear partial differential equations (LPDEs) we introduce an
operator equation for auxiliary operators. These operators are used to
construct a kernel of an integral transformation leading the LPDE to the
separation of variables (SoV).

The auxiliary operators are found for various types of the SoV
including conventional SoV in the scalar second
order LPDE and the SoV by the functional Bethe anzatz.
The operators are shown to relate to separable variables.
This approach is similar to the position-momentum
transformation to action angle coordinates in the classical mechanics.

General statements are illustrated by some examples.
\end{abstract}

\pagebreak
\section{Introduction}\label{Int}

We studied certain possibilities to transform  a linear partial
differential equation (LPDE) into the form allowing the separation
of variables (SoV) by integral transformations.
This 'non-coordinate' SoV differs from the conventional SoV
realized by a purely coordinate transformation.
In the classical mechanics, this approach is similar to the
position-momentum canonical transformations in the phase space
reducing a Hamilton system to the action-angle variables.

One of the realization of  the 'non-coordinate' SoV was found
in the form of the functional Bethe-anzatz (FBA)\cite{Skl90}
appeared in the frame of the $R$- matrix formalism.
The $R$- matrix formalism was developed on the background of
the quantum inverse scattering method \cite{ISM86}.
The FBA-method generates effectively classes of
integrable systems, however it is difficult to apply the FBA to
a certain equation as we believe.

In this paper we consider a relationship between the FBA-
constructions and the conventional method of SoV,
that, in particular, highlights some
ways of the 'direct' non-coordinate SoV, using no
$R$ - matrix formalism constructions.

In the general sense, the  method of separation of variables
in a linear
partial differential equation with $n$ arguments
$(x)=(x_i) _{1\ge i\ge n}$
$\in R^n$ is a procedure which brings the LPDE
\begin{equation}\label{01}
F\psi (x) =0
\end{equation}
into the form where the solution $\psi$ can be written
\begin{equation}\label{02}
\psi _\la (x) =S(x)\Pi _{i=1}^{n}\psi _i (x_i,\la).
\end{equation}
(We discuss the local aspects of the SoV, therefore all the functions
are suggested to be smooth). Here $S(x)$ is a function of $(x)$,
and the function $\psi_i (x_i,\la)$ depends only on $x_i$ and
separation parameters
$\la\in C^{ (n-1)}$. The number of parameters $\lambda$ provides the
'local completeness' of the solution family  (\ref{02}), so, that an
arbitrary solution of (\ref{1}) can be formally expanded into
(\ref{02}).

Below herein we notice some aspects of the theory, that are necessary
for this paper.

The theory of complete and partial separation of variables  in a
scalar LPDE of the second order of both non-parabaloic and parabaloic
types is developed entirely. The theorem on necessary and sufficient
conditions of the SoV has been proved in \cite{ShapVN80}, where  two
basic problems in the SoV theory are solved:
\ben

\item The general
form  of an LPDE of both parabaloic and non-parabaloic types, admitting
SoV  in a system of separable (privileged) coordinates was found. The
correspondent form of the system of ordinary differential equations on
the functions $\psi_i(x_i,\la) $ (separated system) and the SoV
procedure  were presented.

\item An invariant criterium of the SoV was obtained
for a certain equation in an arbitrary
coordinate system. This criterium provides the algorithm of the
transformation from an arbitrary coordinate system to the privileged
system.
\een
The criterium is based on the  existence of a complete set of
pairvise commuting symmetry operators of the first and the second
orders  which  satisfy the special algebraic conditions \cite{ShapVN80}.

The invariant criterium is important because
the SoV is not invariant to a coordinate transformation.

The theorems in \cite{ShapVN80} assume that the separation
of variables in an LPDE of the second order (\ref{01})
can be made by the following transformations:
\begin{itemize}
\item[] a coordinate transformation  of the form $x'=x' (x) $
independent on the separation parameters $\la$,

\item[] a function of the form $\psi' (x)=S (x)\psi ( x) $
where $S(x)$ is a function of $x$.
\end{itemize}
This kind of the SoV we shall call the CSoV (conventional SoV).

Rather different approach to the integrability problem has provided
the discovery of the quantum inverse scatting method \cite{ISM86} which
initiated afterwards  the method of the $R$- matrix. In many
papers (see, for example, \cite{Skl85}, \cite{KomKuz87}, \cite{Skl89},
\cite{KomKuz90}),  the relationship between the $R$ - matrix
formalism and the SoV was found. The approach based on this relationship
and named the FBA was suggested in the paper \cite{Skl90}.
The FBA applied to the classical Hamilton system
admits  a possibility to use a 'non-coordinate'
position-momentum canonical transformation to the privileged variables
unlike  the CSoV method.   In terms of  the
differential equation, the FBA leads to the SoV after some
integral transformation of the LPDE system.
In this approach,  integro-
differential operators can play the role of the
privileged coordinates $x$. In this coordinates, the solution  can be
presented in the multiplicative form (\ref{02}).  The equations on
the functions $\psi _i$ in (\ref{02}) (the separated system) are
non-local in general case.

The study of a quantum system (LPDE system)
can be carried out basing on the FBA by the following scheme:
\ben
\item to find $R$- matrix as a solution of quantum  Yang-Baxter equation,
\item to look for the representation of $R$- matrix algebra
\cite{Skl90} in the class of $(n\times n) -L$  operator matrices.
\item to find a system of LPDEs described
by this operator matrices.
\een
The operator matrix $L$ satisfying all the conditions of the FBA
and describing some LPDE solves the
problem of the SoV:
it generates the form of a privileged coordinates
and the form of separated equations.
As we believe, the direct way to build up all elements of
$R$- matrix formalism for a certain quantum system (a system of LPDEs)
is not found in general case.

In this paper we consider the common features of the
CSoV and the FBA. An approach based on auxiliary operators is proposed.
It leads a certain LPDE to the
SoV by integral transformations. The correspondence of
auxiliary operators to the CSoV and FBA is discussed.
General statements are illustrated by several examples including the
well known quantum Goryachev-Chaplygin top.

\section{Operator equation and its relationship
to the CSoV and FBA} \label{OE}

We consider a quantum system, which is reduced to a
pair of compatible two-dimensional LPDEs of an arbitrary order:
\begin{equation}\label{1}
\begin{array}{ll}
&H_1 (x_1, x_2,\pa_{x_1},\pa_{x_2})\psi (x_1, x_2)=
\epsilon_1\psi (x_1, x_2), \\
&H_2 (x_1, x_2,\pa_{x_1},\pa_{x_2})\psi (x_1, x_2)=
\epsilon_2\psi (x_1, x_2),\\
&[H_1, H_2] =0.
\end{array}
\end{equation}
Here and below $\pa_{x_k}=\displaystyle\frac{\pa}{\pa_{x_k}},\;
\pa_{x_kx_l}=\displaystyle\frac{\pa^2}{\pa_{x_k}\pa_{x_l}},
\;k,l,s,\dots=1,2,\;\epsilon_k$ - are parameters.  Let us define the
linear partial differential operators $U_1$ and $U_2$ of the orders $n_1$
and $n_2$ respectively ($n=0$ corresponds to  the multiplication operator
on a function), by the system:

\begin{equation}\label{2}
\begin{array}{ll}
&[U_1, H_1]+[U_2, H_2] =0,\\
&[U_1, U_2] =0.
\end{array}
\end{equation}
If $U_1$ and $U_2$ are supposed to be  functions,
we have the following

\begin{teo}
If the operators $H_1$ and $H_2$  of the second order
permit the separation of variables in Eqs.(\ref{1}) via CSoV,
then the solution of  Eqs.(\ref{2}) in a class
of functions ($n_1=0, n_2=0$) exists.
\end{teo}

{\it Proof}:
Let  the variables in Eqs.(\ref{1})  are separated in a certain
coordinates $x$. Then the matrices $h_1^{kl}$ and $h_2^{kl}$ of the
coefficients at the second order partial derivatives  of the operators
$H_1$ and $H_2$ have the form:
\begin{equation}\label{3}
h_s^{kl}=\sum_r\Phi^{-1}_{sr}\delta_r^k \delta_r^l,
\end{equation}
where  $\delta_k^l$ is the Kronecker delta and
$$
\Phi_{sq}=\left
(\begin{array}{c c}
f_1 & g_1\\f_2 & g_2\\
\end{array}\right)
$$
is the St\"ackel matrix.
Lower index at the functions $f_s$ and $g_s$ indicates
the number of correspondent privileged variables on which the
function depends. Notice, that a matrix of the form $\Phi
_{ks}=\Phi _{ks}(x_k)$ is  the  St\"ackel matrix by the definition.

For the  operators $H_1$ and $H_2$ of the second order,
Eqs. (\ref{1}) takes the form
$$
\sum_l (h_1^{kl}U_{1, l}+h_2^{kl}U_{2, l})=0,
$$
($U_{s,l} = \pa U_s/\pa x_l$) which,  taking into the account
Eq.(\ref{3}) in the privileged variables $V_1$ and $V_2$, results in
\begin{equation}\label{5}
\Phi^{-1}_{1s}U_{1, s}+\Phi^{-1}_{2s}U_{2, s}=0.
\end{equation}
The integration of Eq.(\ref{5}) gives:
$$
U_k=\sum_l\Phi^{-1}_{kl}q_\l,
$$
where $q_1,\;q_2$ are arbitrary functions of
$V_1$ and $V_2$, respectively.

The relationship between the functions $U_1$ and $U_2$ and the
separable variables $V_1$ and $V_2$ can be obtained from the
following
\begin{teo} If  the operators of the second order $H_1$ and
$H_2$ satisfy  the CSoV conditions,  and  $V_1$ and $V_2$ are the
privileged coordinates then Eq. (\ref{2}) is fulfilled for
$U_1=V_1+V_2$ and $U_2=V_1V_2$.  \end{teo}

{\it Proof}: According to the CSoV,
the  privileged coordinates  can be taken in the form:
\begin{equation}\label{7}
\begin{array}{ll}
&V_1=1/2 (\tr (h_1\cdot h_2^{-1})+\sqrt{\tr (h_1\cdot h_2^{-1})
^2+4\det (h_1\cdot h_2^{-1}) }), \\
&V_1=1/2 (\tr (h_1\cdot h_2^{-1}) -
\sqrt{\tr (h_1\cdot h_2^{-1}) ^2+4\det (h_1\cdot h_2^{-1}) }).
\end{array}
\end{equation}
Let us show, that for $U_1$ and $U_2$ we can take
\begin{equation}\label{8}
\begin{array}{ll}
&U_1=V_1+V_2=\tr (h_1\cdot h_2^{-1}),\\
&U_2=V_1\cdot V_2=\det (h_1\cdot h_2^{-1}).
\end{array}
\end{equation}
Substitution of the matrices $h_1$ and $h_2$ of the form (\ref{3})  into
the expressions for the trace and the determinant yields:

\begin{equation}\label{9}
\begin{array}{ll}
&U_1=\tr (h_1\cdot h_2^{-1})=(f_2g_1+f_1g_2)/{f_1f_2},\\
&U_2=\det (h_1\cdot h_2^{-1}) =g_1g_2/f_1f_2.
\end{array}
\end{equation}
From (\ref{5}) it follows
\begin{equation}\label{10}
\begin{array}{ll}
&(g_2U_1-f_2U_2), _1=0,\\
&(g_1U_1-f_1U_2), _2=0.
\end{array}
\end{equation}
Substitution of $U_1$ and $U_2$ from (\ref{9}) into (\ref{10}) gives:
$$ (g_2^2/f_2),_1=0,\;\; (g_1^2/f_1), _2=0. $$

The similar connection between
the FBA and the system (\ref{2}) exists if  $U_1$ and $U_2$
are operators.
The separated system can  be written after the transformation to
the representation of these operators.

We consider the quantum system described by
($2\times 2$) $L$- matrix associated with yangian $Y[gl(2)]$:
$$L(u)=\left (
\begin{array}{c c}
A (u) & B (u)\\C (u) & D (u)\\
\end{array}
\right). $$
Here $A(u),\;B(u),\;C(u)$ and $D(u)$ are linear differential operators
depending on the spectral parameter $u$.
Following to the $R-$ matrix formalism \cite{Skl90} we have:
\begin{equation}\label{11}
\tr(L (u)) =A (u) +D (u) =u\cdot H_2+H_1
\end{equation}
with
\begin{equation}\label{12}
\begin{array}{ll}
&[A (V_k), V_l]=\mu A (V_k)\delta_{kl}, \\
&[D (V_k), V_l]=-\mu D (V_k)\delta_{kl}.
\end{array}
\end{equation}
Here $\delta_{kl}$ is the Kronecker delta and $\mu$ is the parameter of
$R$ - matrix.  According to the FBA, operators  $V_s$ become the separable
variables after transformation to their representation.
From commutators for $\tr(L(V_k))$ with separable variables $V_l$ and
Eqs.(\ref{11}), (\ref{12}),
we obtain:
\begin{equation}\label{15}
[H_1, V_1+V_2]+[H_2, V_1\cdot V_2] =0.
\end{equation}
So, it is true the following:
\begin{teo}
If the system (\ref{1}) permits SoV via the FBA
then the operators $U_1$ and $U_2$ satisfying (\ref{2}) exist, where
$U_1=V_1+V_2$ and $U_2=V_1V_2$. Here $V_1$ and $V_2$ are the operators
correspondent to the separable by the FBA variables.
\end{teo}

The existence of the operators  $U_1$ and $U_2$  satisfying
(\ref{2}) can be understood as a necessary condition for the
applicability of the FBA method.

Let us consider a special case of Eq.(\ref{2}) where
each term in it equals to zero:
\begin{equation}\label{50}
[H_1, U_1]=[H_2, U_2]=0.
\end{equation}
Eq. (\ref{50}) appears  in the  simplest cases of the SoV.
Eq.(\ref{50}) have the solution in the class of functions if
each of the equations in the system (\ref{1}) includes only one variable
or if the SoV in (\ref{1}) can be obtained by the transformations of
variables without linear combination of the equations. If $H_1$ and
$H_2$ in (\ref{1}) are of the form
$H_1=H_1(U_1,U_2)$ and
$H_2=H_2(U_1,U_2)$ then this $U_1$ and $U_2$ are
the solution of (\ref{50}).

We showed that the various types of the SoV relate to the auxiliary
operators, which are the solutions of (\ref{2}) of
the correspondent type. This idea is investigated
in the approach below.

\section{Integral transformations} \label{I}

Let us consider an integral transformation  of the system
(\ref{1}) with kernel $K (x_1, x_2;\la_1,\la_2) $ defined
by the system of differential equations:
\begin{equation}\label{17}
\begin{array}{ll}
&U_1 (x_1, x_2,\pa _{x_1},\pa _{x_2}) K (x_1, x_2;\la _1,\la _2)=
\la_1K (x_1, x_2;\la _1,\la _2),  \\
&U_2 (x_1, x_2,\pa _{x_1},\pa _{x_2}) K (x_1, x_2;\la _1,
\la _2)=\la _2K (x_1, x_2;\la _1,\la _2).
\end{array}
\end{equation}
Here $U_1$ and $U_2$ are the solutions of (\ref{2})

Let us take $\psi (x_1,x_2) $ in the form:
\begin{equation}
\label{18}
\psi (x_1, x_2)=\int K (x_1, x_2;\la _1,\la _2)
\phi (\la _1,\la _2) d\la _1 d\la _2.
\end{equation}
Inserting (\ref{18}) into  (\ref{1}) and converting the action
of the operators $H_1$ and $H_2$ from $\psi (x_1, x_2)$ to
$\phi (\la _1,\la _2) $ we obtain:
\begin{equation}\label{19}
\begin{array}{ll}
&R_1 (\la_1,\la_2,\pa_{\la_1},\pa _{\la _2})\phi (\la _1,\la _2) =
\epsilon_1\phi (\la_1,\la_2),\\
&R_2 (\la _1,\la _2,\pa _{\la _1},\pa _{\la _2})\phi (\la _1,\la_2) =
\epsilon_2\phi (\la _1,\la _2).
\end{array}
\end{equation}
Here $R_1$ and $R_2$ are integro- differential operators in the
general case.  The variables can be separated in the system (\ref{19})
under the proper choice of $U_1$ and $U_2$.
The theorems 1 and 3 of the previous section serve for
the arguments in this approach.

As mentioned above, $U_1$ and $U_2$ are the functions
in the CSoV  case. Then the solution of Eq. (\ref{17}) is  the
Dirak delta. The transformation to the system (\ref{19}) is identical.

If $U_1$ and $U_2$ are the operators, the integral transformation
can lead to SoV the systems, that are not separable by a coordinate
transformation.  These examples are discussed in the next section.
The problem of the proposed approach is the complexity of
the transformation to the system (\ref{19}), if the kernel (\ref{17})
is complex, and the proper choice of $U_1$ and $U_2$ from all
solutions of (\ref{2}).

Notice, that after transformation to the suitable
$(U_1,\; U_2)$ representation, the separable variables $V_1$
and $V_2$ can be found from: $U_1=V_1+V_2,\; U_2=V_1V_2$.

\section{Examples}
\subsection{The system of LPDEs of the second order
not permitting CSoV}

If the system (\ref{1}) permits the CSoV, then
the operators $H_1$ and $H_2$ must satisfy some
algebraic conditions \cite{ShapVN80}. In the  example below,
this algebraic condition is broken,
but it recovers after integral transformation  according
the approach  of sec. \ref{I}.
Let the operators $H_1$ and $H_2$ in (\ref{1}) are:
\begin{equation}\label{20}
\begin{array}{ll}
&H_1=\pa_{x_1x_1}+\pa _{x_2}+\mu (x_2^2\pa _{x_2}+x_1^2), \\
&H_2=x_2\pa _{x_1x_2} -\frac{1}{2}x_1\pa _{x_2}+\frac{1}{2}\mu
x_1x_2^2\pa_{x_2},
\end{array}
\end{equation}
where $\mu$ is a constant.
The system does not satisfy
the necessary condition of the CSoV, which can be formulated here in the
form:
$$ \tr (h_1\cdot h_2^{-1}) ^2+4\det (h_1\cdot h_2^{-1}) > 0. $$
Here $h_1$ and $h_2$ are the matrices of the coefficients at the second
order partial derivatives of the operators $H_1$ and $H_2$ respectively.
Solving the system (\ref{2}) we obtain, in particular:
$$
\begin{array}{ll}
&U_1=x_1,\\
&U_2=x_2^2\pa _{x_2}.
\end{array}
$$
The solution of (\ref{17}) is $K=\exp{ (-\la_2/x_2)}\delta (x_1 -\la_1)$.
The integral transformation yields to:
\begin{equation}\label{21}
\begin{array}{ll}
&R_1=\pa _{\la _1\la _1}+\la _2\pa _{\la _2\la _2}+\mu (\la _1^2+\la
_2^2),\\
&R_2=\la _2\pa_{\la _1\la _2} -\frac{1}{2}\la _1\la _2\pa
_{\la _2\la _2}+ \frac{1}{2}\mu\la _1\la _2.  \end{array} \end{equation}

The system (\ref{21}) allows the CSoV:
$$
\begin{array}{ll}
&T (V_1)\psi (V_1) =0,\\
&T (V_2)\psi (V_2) =0,\\
&T (V)=\pa_{VV}+\mu/4V^2-\epsilon_1-\epsilon_2/V.
\end{array}
$$
Here the separable variables $V_k$ are related to $U_k$ as follows:
$$
\begin{array}{ll}
&V_1=U_1+U_2,\\
&V_2=U_1\cdot U_2.
\end{array}
$$

\subsection{Quantum Goryachev-Chaplygin top}

In this example we apply the approach to the well known quantum
system studied by the FBA method in \cite{KomKuz87}.

Let us consider the quantum analogue
of the Hamiltonian system on the Lie algebra $e (3) $.
The generators can be taken in the class of  differential operators
in the space $R^3$ with the conventional commutators:

\begin{equation}
[J_i, J_j]=-i\sum_k\epsilon_{ijk} J_k,\;\;\; [J_i,
x_j]=-i\sum_k\epsilon_{ijk} x_k,\;\;\; [x_i, x_j] =0.
\end{equation}
Here $i,j,k=1,2,3$.

The Casimir operators have the  form:
\begin{equation}\label{002}
K_1=x_1^2+x_2^2+x_3^2,\;\;\;
K_2=x_1 J_1+x_2 J_2+x_3 J_3.
\end{equation}
The Quantum Goryachev-Chaplygin top
is described  by the Hamiltonian $H$ and the additional integral
of motion $G$\cite{KomKuz87}:

\begin{equation}\label{003}
\begin{array}{ll}
&H=\frac{1}{2} (J_1^2+J_2^2+4J_3^2)-bx_1, \\
&G=(2J_3+p)(J_1^2+J_2^2 -\frac{1}{4}) +b [x_3, J_1] _+ .
\end{array}
\end{equation}
Here $b$ is a constants, $[A,B]_+=AB+BA$.
The operators $H$ and $G$ commute on the subspace of  eigenfunctions
$\psi$ of the Casimir operators matching the eigenstates equal to $1,\;0$:
\begin{equation}\label{33}
K_1(J, x)\psi=1\cdot \psi,\;\;\; K_2(J, x)\psi=0.
\end{equation}
In this case the system of Eq.(\ref{002}) and
(\ref{003}) is integrable.

The reduction of $H$ and $G$ on the selected integrable orbit
is obtained here by the method of noncommutative
integration of the LPDE developed in \cite{Sh95}.
Written in the variables of the orbit $\zeta_1,\zeta_2$,
the system has a form:

\begin{equation}\label{443}
\begin{array}{ll}
&H=\frac{1}{2} ((4+\tan{\zeta _2}^2)\partial_{\zeta _1\zeta_1}-
\tan{\zeta _2}\partial_{\zeta _2}+\partial_{\zeta_2\zeta_2})-
b\sin{\zeta _1}\cos{\zeta _2}, \\
&G=2\partial _{\zeta _1} (\tan{\zeta _2}^2\partial_{\zeta _1\zeta_1}+
\partial_{\zeta_2\zeta_2} -\tan{\zeta _2}\partial _{\zeta _2}-1/4)-\\
&b(2\sin{\zeta _1}\sin{\zeta _2}\tan{\zeta _2}\partial _{\zeta _1}+
2\cos{\zeta _1}\sin{\zeta _2}\partial _{\zeta _2}+
\cos{\zeta _1}\cos{\zeta _2}).
\end{array}
\end{equation}

This system can be considered as the quantum
analogue of the reduced system on the orbit of the
coadjoint representation of the Lie algebra $e(3)$.

In this system the constraints (\ref{33}) produced by the
Casimir operators are eliminated  and the condition $[H, G]$=0 is
satisfied.

Solving system (\ref{2}) for $H_1=H$ and $H_2=G$ we
obtain,  in particular:
\begin{equation}
U_1=\tan{\zeta _2}^2\pa_{\zeta_1\zeta_1}-
\tan{\zeta_2}\partial_{\zeta _2}+\partial_{\zeta _2\zeta_2},\;\;
U_2=2\partial_{\zeta _1}.
\end{equation}
According to our approach the separable variables $V_1$ and $V_2$
are related to $U_1$ and $U_2$  as follows:
\begin{equation}\label{444}
U_1=V_1+V_2,\;\;U_2=V_1\cdot V_2.
\end{equation}
This result can be interpreted in the frame of the FBA.
The separable variables are by the FBA of the form \cite{KomKuz87}:

\begin{equation}\label{446}
\begin{array}{ll}
&V_1=J_3+\sqrt{J_1^2+J_2^2+J_3^2},\\
&V_2=J_3 -\sqrt{J_1^2+J_2^2+J_3^2}.
\end{array}
\end{equation}
After the substitution of (\ref{446}) in (\ref{444})
in the orbit variables $\la _1,\la _2$ we
obtain (\ref{443}), which is defined from our approach.

The system (\ref{19}) is difficult to derive
due to the nonanalytic kernel of integral transformation
defined by (\ref{17}).

It is known from the FBA, that the system (\ref{003}) in terms of the
variables $V_1$ and $V_2$ takes the separable form:
$$
\begin{array}{ll}\label{555}
&T (V_1)\psi (V_1) =0,\\
&T (V_2)\psi (V_2)=0,\\
&T (V) =V^3-2 (h+1/8) V+g-b \sqrt{ (V-1) ^2+1/4}
e^{2\pa_V}-\\
&b \sqrt{ (V+1) ^2+1/4}e^{-2\partial_V}.
\end{array}
$$
Here $f$ and $g$ are the eigenfunctions of $H$ and $G$, respectively.
In (\ref{555}), $V_1$ and $V_2$ can be taken from (\ref{444}).  We can
interpret the separable variables as the quantum analogue
variables of the integrable coadjoint orbit of the of the Lie algebra
$e(3)$.

\subsection{Electron dynamics in resonant cyclic accelerator}

We apply our approach to  a quantum system,
which exact solution was found in \cite{Bagr70}.
The quantum system is determined by the Hamiltonian:
\begin{equation}\label{60}
\begin{array}{ll}
&H_1=\pa_ {xx} -
2i\gamma x\pa_{\xi}-a^2x^2
-\gamma \gamma _0\xi^2 \\
&-a_3^2\displaystyle\frac{z^2}{f(z)}+f(z)\pa_{zz}.
\end{array}
\end{equation}
The parameters $a_3,\gamma,\gamma_0$ of the Hamiltonian  have a
definite physical sense, $f(z)$ is a function.  The variable $z$ can be
simply separated. The symmetry operator of the second order can be found
as:
\begin{equation}\label{61}
H_2=-2i\gamma \gamma_0
\xi\pa_x+ \gamma _0\xi^2 -\gamma ^2\gamma _0x^2
-\gamma \pa_{\xi\xi}.
\end{equation}
This system does not satisfy the CSoV conditions.
One of the solutions of the equation (\ref{2}) is:
\begin{equation}\label{62}
U_1=\pa_{\xi},\quad U_2=x.
\end{equation}
Substituting (\ref{62}) in (\ref{17}) we obtain, with respect to $x_1=x$
and $x_2=\xi$, identical transformation for the  variable $x$ and Fourier
transformation for $\xi$. This is similar to the results
of \cite{Bagr70}.  After this transformation, the system (\ref{60}),
(\ref{61}) takes the form:
\begin{equation}\label{63}
\begin{array}{ll}
&R_1=\pa_{\la_1\la_1} -
\gamma \gamma _0\pa_{\la_2\la_2}+ 2\gamma
\la_1\la_2-a^2 \la_1^2,\\
&R_2= -\gamma _0\pa_{\la_2\la_2}+
2\gamma \gamma _0\pa_{\la_1\la_2} -
\gamma ^2\gamma _0 \la_1^2+\gamma \la_2^2.
\end{array}
\end{equation}

The transformation of variables
$$
\begin{array}{ll}
&V_1=\displaystyle\frac{\gamma }{a_4 \sqrt{a_1}}
(\displaystyle\frac{a_1^2}{\gamma }\lambda_1-\lambda_2), \\
&V_2=\displaystyle\frac{\gamma }{a_4 \sqrt{a_2}}
(\lambda_2 -\displaystyle\frac{a_2^2}{\gamma }\lambda_1)
\end{array}
$$
Here $a_1,a_2,a_4$ are expressed in terms of
$\gamma_0,\gamma,a$.  This transformation reduces
the system (\ref{63}) to the separated form:
$$
\begin{array}{ll}
&(\pa_{V_1V_1} -V_1^2)\psi(V_1,V_2)=\epsilon_1\psi (V_1,V_2), \\
&(\pa_{V_2V_2} -V_2^2)\psi(V_1,V_2)=\epsilon_2\psi (V_1,V_2).
\end{array} $$

\section{Conclusion}

We showed the interrelation between  the
separable variables of the FBA and the method of complete
SoV in the scalar second order equation in terms of auxiliary
operators $U_1$ and $U_2$. The conventional SoV appeared to be
the simplest type, if $U_1$ and $U_2$ are operators of
multiplication on the functions. The FBA theory, which generalizes the
theory of SoV, produces the operators $U_1$ and $U_2$ in more complicate
form.

We put in the background of
our approach the operators $U_1$ and $U_2$, which are the solutions of
Eq.(\ref{2}) in class of the differential operators.
It provides a 'direct' way of the
'non-coordinate' SoV for an LPDE system.
It can be applied, for example, to the systems with higher
order symmetries.
Such a problem was researched in various papers (see, for example
\cite{Hiet87}).

It is interesting, from our point of view,
to study the following items:
\ben
\item[] the sufficient condition of the SoV in terms
of the auxiliary operators
\item[]  the analogue of Eq.(\ref{2}) in 3
and more dimensions can be the theme of latest development
\item[] the
structure of the auxiliary operators for the different types of LPDE.
\een
It will be the subject of the further papers.


\begin{thebibliography}{99}
\bibitem{Skl90} Sklyanin E. K. In: Integrable and Superintegrable
Systems (Ed. by B. A.  Kupershmidt, Singapore: World
Scientific) (1990) p. 8-33.
\bibitem{ISM86} Takhtajan L. A., Faddeev L. D.: A Hamiltonian approach
in the soliton theory. Moscow, Nauka (1986) (in Russian).
\bibitem {ShapVN80} Shapovalov V. N. Differential
equations {\bf 16} (1980) p. 1864-1874. (in Russian).
\bibitem{Skl85}  Sklynin E. K.
Journ. Sov. Math. 31 (1985) p. 3417-3431. (in Russian).
\bibitem{Skl89} Sklynin E. K., J. Sov. Math.  46 (1989) p. 1664-1683.
\bibitem{KomKuz87} Komarov I. V., Kuznetsov V. B. 1987. Zap. Nauch. Semin.
LOMI {\bf 164} p. 134-142. (in Russian).
\bibitem{KomKuz90} Komarov I. V., Kuznetsov V. B. J. Phys. A: Math.Gen.
(1990) {\bf 23} p. 841-846.
\bibitem{Sh95} Shapovalov A. V., Shirokov I. V.
Theor. Math. Phys.  {\bf 104} (1995) p. 195-213. (in Russian).
\bibitem{Bagr70}  Sokolov A. A.,  Ternov I. M.,  Bagrov V. G.
Annalen der Physik  25 (1970) p. 44-49.
\bibitem{Hiet87} Hietarinta J.  Phys. Rep. {\bf 147} (1987) p. 87-154.
\end{thebibliography}
\end{document}